\documentclass[pra,preprint,superscriptaddress,floatfix,showpacs, 11pt]{revtex4-1}

\usepackage{graphicx}
\usepackage{dcolumn}
\usepackage{bm}
\usepackage{xcolor}
\usepackage{amsmath}
\usepackage{hyperref}
\usepackage[normalem]{ulem}

\begin{document}


\title{Rotation-mediated bosonic Josephson junctions in position and momentum spaces}
\author{Sunayana Dutta}
\email{sdutta@campus.haifa.ac.il}
\affiliation{Department of Physics, University of Haifa, Haifa 3498838, Israel}
\affiliation{Haifa Research Center for Theoretical Physics and Astrophysics, University of Haifa, Haifa 3498838, Israel}
\author{Rhombik Roy}
\affiliation{Department of Physics, University of Haifa, Haifa 3498838, Israel}
\affiliation{Haifa Research Center for Theoretical Physics and Astrophysics, University of Haifa, Haifa 3498838, Israel}
\author{Ofir E. Alon}
\affiliation{Department of Physics, University of Haifa, Haifa 3498838, Israel}
\affiliation{Haifa Research Center for Theoretical Physics and Astrophysics, University of Haifa, Haifa 3498838, Israel}


\begin{abstract}
     In ultracold atoms, bosons tunneling in a double-well potential can produce a typical Josephson junction in real space. A major advancement in quantum matter and simulations is anticipated by the recently found momentum-space Josephson junctions, which elucidates the supercurrent flow between spin-orbit coupled Bose-Einstein condensates at two distinct independent momentum states. For the first time, our study unveils specific protocols to engineer momentum-space Josephson dynamics for scalar bosons (or a single-component condensate) in a rotating frame through modulation of the geometry of a double-well trapping potential and rotation frequencies. In this setup, the rotation simultaneously results in effective double wells both in position and in momentum spaces, and the dynamics of the corresponding Josephson junctions is hosted in these double wells. Consequently, it is observed that the rotation generates momentum-space Josephson dynamics of the condensate along the transverse direction and position-space Josephson dynamics along the longitudinal direction; these effects are particularly noticeable for high rotation frequencies. Additionally, the rotation-induced momentum-space junctions are highly significant in both the mean-field and many-body dynamics. Our protocols offer a framework for investigating momentum-space Josephson junctions for single-component condensates in both theoretical and experimental contexts, as well as their significant applications in quantum mechanical devices.
\end{abstract}

\maketitle                             

\section{Introduction}\label{introduction}
The Josephson effect \cite{Josephson1962possible, josephson1974discovery} is a fascinating quantum phenomenon in which supercurrents flow through two reservoirs coupled by a weak link known as a Josephson junction. On numerous platforms, the Josephson junction has been experimentally realized, extending from solid-state superconductors \cite{anderson1963probable}, superfluid Helium \cite{backhaus1997direct,wheatley1975experimental,leggett1975theoretical,sukhatme2001observation,hoskinson2005quantum}, exciton polaritons \cite{abbarchi2013macroscopic} to ultracold atoms \cite{dalfovo1996order,andrews1997observation,smerzi1997quantum,ohberg1999internal,williams1999nonlinear,raghavan1999coherent,albiez2005direct,levy2007ac,zibold2010classical,spagnolli2017crossing,burchianti2018connecting,luick2020ideal,valtolina2015josephson}. The latter case, for bosons tunneling in a double-well potential, is commonly referred to as a bosonic Josephson junction (BJJ). Josephson junctions play a crucial role in the development of superconducting quantum interference devices (SQUIDs) \cite{makhlin2001quantum,ryu2013experimental}, superconducting qubits \cite{martinis2002rabi,astafiev2006temperature,martinis2009energy,paik2011observation}, and precision measurements \cite{makhlin2001quantum,rhombik_acc}, due to quantum tunneling of particles across the junction. 
Exploration of BJJs leads to uncover various intriguing features, including Josephson-like dynamics of a BEC of rubidium atoms \cite{vargas2021orbital}, universality of fragmentation dynamics of an asymmetric BJJ \cite{sakmann2014universality}, impact of both the transversal and longitudinal degrees-of-freedom on Josephson dynamics in the resonant tunneling scenario \cite{bhowmik2022longitudinal}, among others. 

In recent times, utilizing momentum states as a synthetic degree of freedom for quantum matter and quantum simulations \cite{an2018correlated} has become more feasible with the realization of the momentum-space Josephson junction. A momentum-space Josephson junction can be realized with a momentum-space double-well dispersion, where the condensates at two distinct band minima can be treated as two distinct independent quantum states. Tunneling then occurs between these two momentum states through their mutual coupling, in contrast to conventional real-space Josephson tunneling between spatially separated wells. The momentum-space Josephson Junction has been first reported for a spin-orbit-coupled BEC located at two discrete momentum states coupled through momentum kicks of laser beams \cite{hou2018momentum}. Beyond the conventional voltage-driven Josephson effect, Ref.~\cite{hou2018momentum} demonstrates how the tunneling phase in momentum space can generate the Josephson junction. 
 Also, in a Raman-induced spin-orbit-coupled BEC, Ref. \cite{mukhopadhyay2024observation} experimentally realized the momentum-space Josephson effect, whose double-well band dispersion has two band minima at distinct momentum states that resemble real-space Josephson junctions. 

Lately, the breakup of a rotating dipolar condensate into two localized fragments, each with an equal number of vortices, was analyzed in three dimensions in Ref.~\cite{kumar2016three}, and the splitting of the condensate into two localized fragments with a giant vortex, for a rotating pancake-like asymmetric quartic-quadratic potential, demonstrated in~\cite{gamma}. Furthermore, we reported the breakup of the ground-state density of rotating bosons in an elongated barrierless trap for finite systems \cite{dutta2023fragmentation,rhombik_scirep} as well as in the limit of an infinite number of particles \cite{dutta2} in previous works. In addition, it is noted that the density fragments into two bosonic clouds in position space along the longitudinal direction and in momentum space in the transverse direction, thereby creating two effective double wells. Hence, the results from these ground-state investigations invoke possible ideas of engineering bosonic Josephson junctions in position and momentum spaces for scalar bosons using rotations. 
In the present work, the starting point of our investigation is the impact of rotation on the Josephson dynamics of scalar bosons in a double-well potential. In the weakly interacting and relatively low-rotation regime considered here, no vortex formation is observed~\cite{Stringari,fetter}. It is intriguing to examine whether rotation can enhance or suppress the Josephson dynamics in the double-well potential. Further, the possibility of generating the momentum-space Josephson junction is another primordial point. Our work is further extended to unveil a useful protocol, based on the double-well potential, to engineer rotation-induced Josephson junctions both in position and momentum spaces for scalar bosons. 

\section{Theoretical Method}\label{theory}
We consider the multi-configurational time-dependent Hartree method for bosons (MCTDHB)  \cite{streltsov:07,alon:08} to numerically simulate 
interacting bosons confined in a double-well potential in the rotating frame. The dynamical properties of these trapped bosons can be described by the time-dependent many-body Schr\"{o}dinger equation. The Schr\"{o}dinger equation that deals with a many-boson system is frequently solved with the help of the mean-field Gross-Pitaevskii equation. However, the Gross-Pitaevskii approximation is unable to capture many-body features such as fragmentation and correlations, owing to its building via the mean-field ansatz, which has only a single basis state. The MCTDHB method uses the (time-dependent) optimized one-body basis (orbitals). The set of time-dependent orbitals and the expansion coefficients in the MCTDHB basis are optimized variationally \cite{streltsov:07,alon:08}. In addition, the MCTDHB method is in principle numerically exact \cite{lode2012numerically} and able to describe both condensed and fragmented bosons. One can say that the theory of Gross-Pitaevskii approximation is a special case of MCTDHB when only a single one-body orbital is in use. The MCTDHB method is well documented in the literature and hence elaborated in the supplemental material in Sec. S1.

The $N$-boson Hamiltonian can be written as
\begin{eqnarray}
\hat{H} = \sum_{i=1}^{N} \hat{h}(\mathbf{r}_i) + \sum_{i<j} \hat{W}(|\mathbf{r}_i-\mathbf{r}_j|). 
\label{eq:hamiltonian}
\end{eqnarray}
Here, $\hat{h}(\mathbf{r})=\hat{T}(\mathbf{r})+\hat{V}(\mathbf{r})$ is the one-body operator. In the rotating frame, the kinetic energy in the one-body operator is modified with the angular-momentum operator $\hat{l}_z$ as $\hat{T}({\bf r })  = \frac{1}{2} (\hat{p}_{x}^2+\hat{p}_{y}^2) + \Omega\hat{l}_z$, with $\Omega$ being the rotation frequency. In addition, $\hat{V}(\mathbf{r})$ represents the trapping potential. In our work, the trapping potential is a two-dimensional double-well potential given as
\begin{eqnarray}
\hat{V}({\mathbf{r}}) &=&  
\begin{cases}
\frac{1}{2}(x+2)^2+\frac{1}{2}y^2; ~x<-2,~ -\infty  <y< \infty\\
\frac{\alpha}{2}(x+2)^2+\frac{1}{2}y^2;~ -2\le x <-\frac{1}{2}, ~ -\infty <y<\infty\\
\frac{3\alpha}{2}(1-x^2)+\frac{1}{2}y^2;~ -\frac{1}{2} \le x\le\frac{1}{2},~ -\infty <y<\infty\\
\frac{\alpha}{2}(x-2)^2+\frac{1}{2}y^2;~ \frac{1}{2}<x \le 2, ~ -\infty <y<\infty\\
\frac{1}{2}(x-2)^2+\frac{1}{2}y^2;~  x>+2,~ -\infty  <y< \infty\ 
\label{eq:one}
\end{cases} ,
\end{eqnarray}
where $\alpha$ scales the barrier height of ${V}(\mathbf{r})$.
Finally, the two-body operator $\hat{W}(|\mathbf{r}-\mathbf{r}'|)=\frac{\lambda_0}{2 \pi \sigma^2} e ^{-\frac{({\bf r}-{\bf r'})^2}{2\sigma^2}}$ is the repulsive interaction between the bosons. The interaction is modeled by a Gaussian potential, as the delta potential does not scatter in two dimensions \cite{doganov2013two,rhombik_jcp}. $\sigma=0.25$ is the interaction width,  $\lambda_0$ is the interaction strength and is scaled with the number of bosons $N$ as $\Lambda=\lambda_0(N-1)$, where $\Lambda$ is called the interaction parameter. 
 For the numerical simulation, the MCTDH-X software \cite{lode:20,lode2019mctdh,lin2020mctdh,sunayana_tutorial}, which implements the MCTDHB theory, is used. In this work, to explore the many-body effects, $N = 10$ interacting bosons with the interaction parameter $\Lambda=0.04$ are used. We also analyze the mean-field dynamics with the same $\Lambda$. The grid used to represent the Hamiltonian [Equation \eqref{eq:hamiltonian}]
extends from $[-10, 10)$ to $[-10, 10)$ and comprises of $64 \times 64$ discrete variable representation (exponential) functions to represent each of the orbitals.
For our many-body calculations, $M=6$ self-consistent orbitals are used. The convergence of the results presented here, with respect to the number of orbitals, system size, and grid density, is discussed in detail in the supplemental material.

For the investigation of the dynamics under rotation,
the bosons are initially prepared in the left well $V_L(x,y)=\frac{1}{2}(x+2)^2+\frac{1}{2}y^2$ of a double-well potential. The potential is suddenly quenched from $V_L$ to $V$ and the out-of-equilibrium dynamics is followed.

To adequately capture the essence of Josephson dynamics in a double well [see Equation \eqref{eq:one}], the concept of survival probability in the left well is used. The survival probability signifies the degree of tunneling of the bosons from the left to the right wells,
\begin{eqnarray}
 P_{L}(t)=\int_{x=-\infty}^{x=0}\int_{y=-\infty}^{y=+\infty} dx dy \frac{\rho(x,y;t)}{N}, 
\end{eqnarray}
 with $\rho(x,y;t)$
  being the density of the bosonic cloud at the instant of time $t$. 

All the results in this paper are presented with a modified time. The propagation time is scaled by the time period of the dynamics, which corresponds to Rabi oscillations of non-interacting bosons in the stationary double-well potential, $t_{Rabi}$; here $t_{Rabi}=132.498$.  
\begin{figure*}[htbp]
 \centering 
\includegraphics[width=0.24\textwidth]{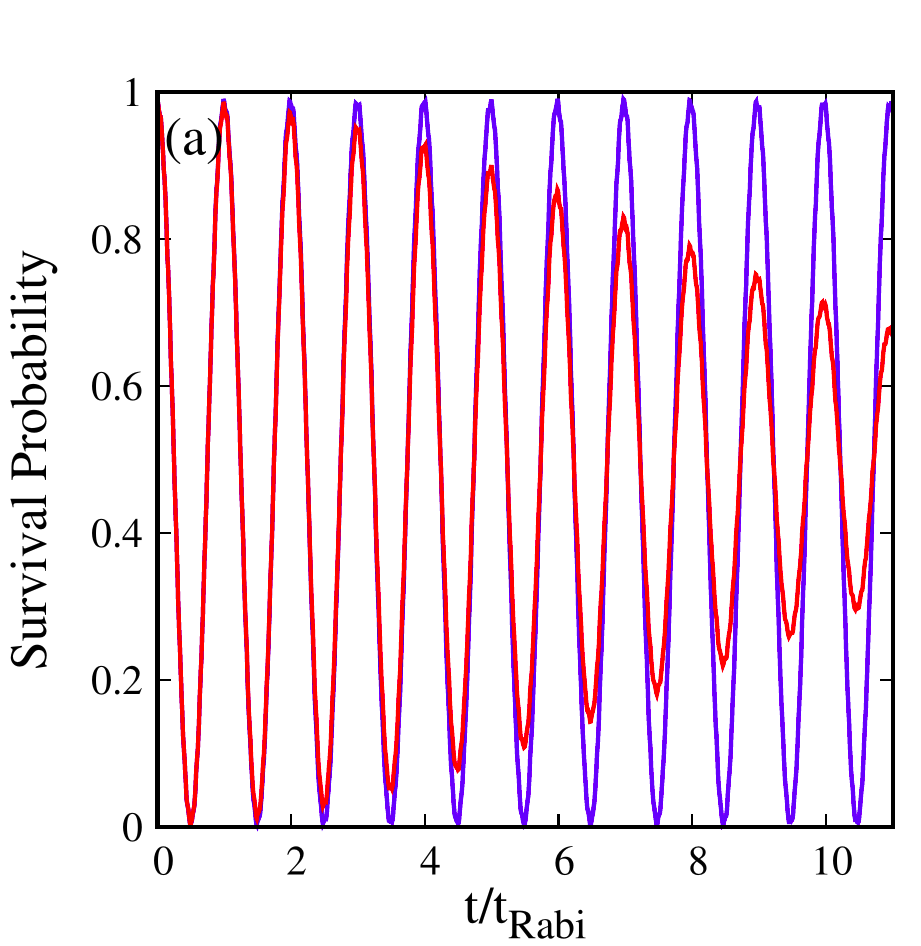}
\includegraphics[width=0.24\textwidth]{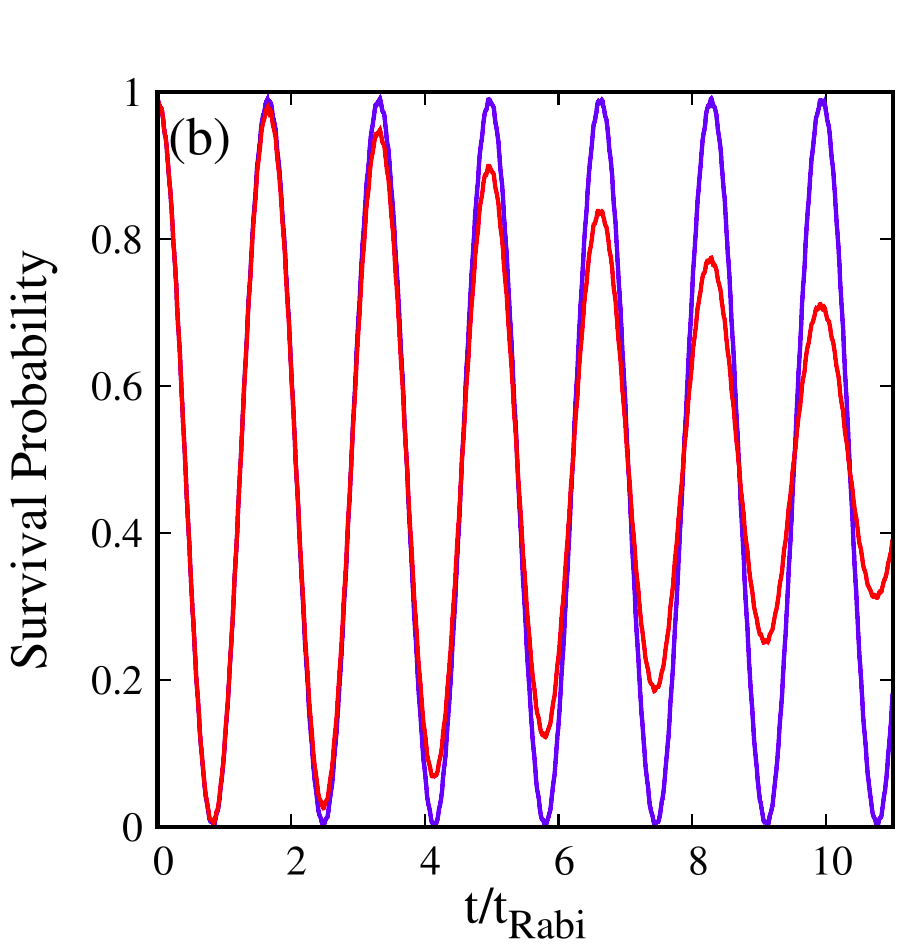}
\includegraphics[width=0.24\textwidth]{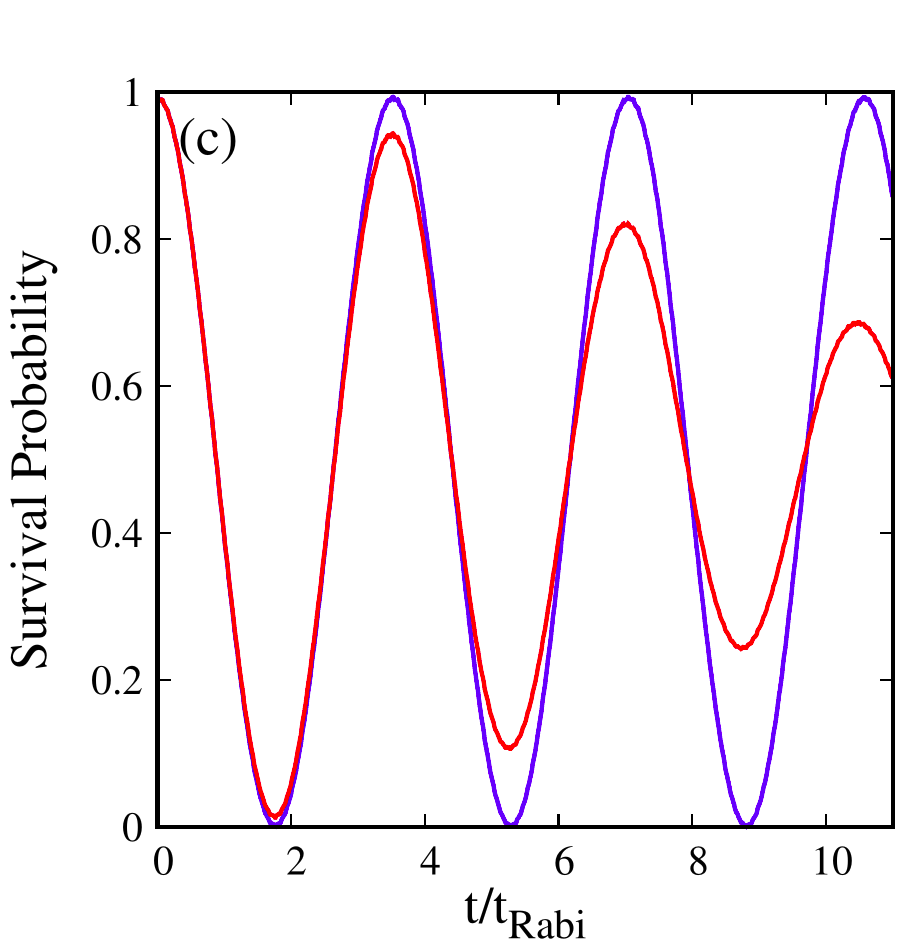}
\includegraphics[width=0.24\textwidth]{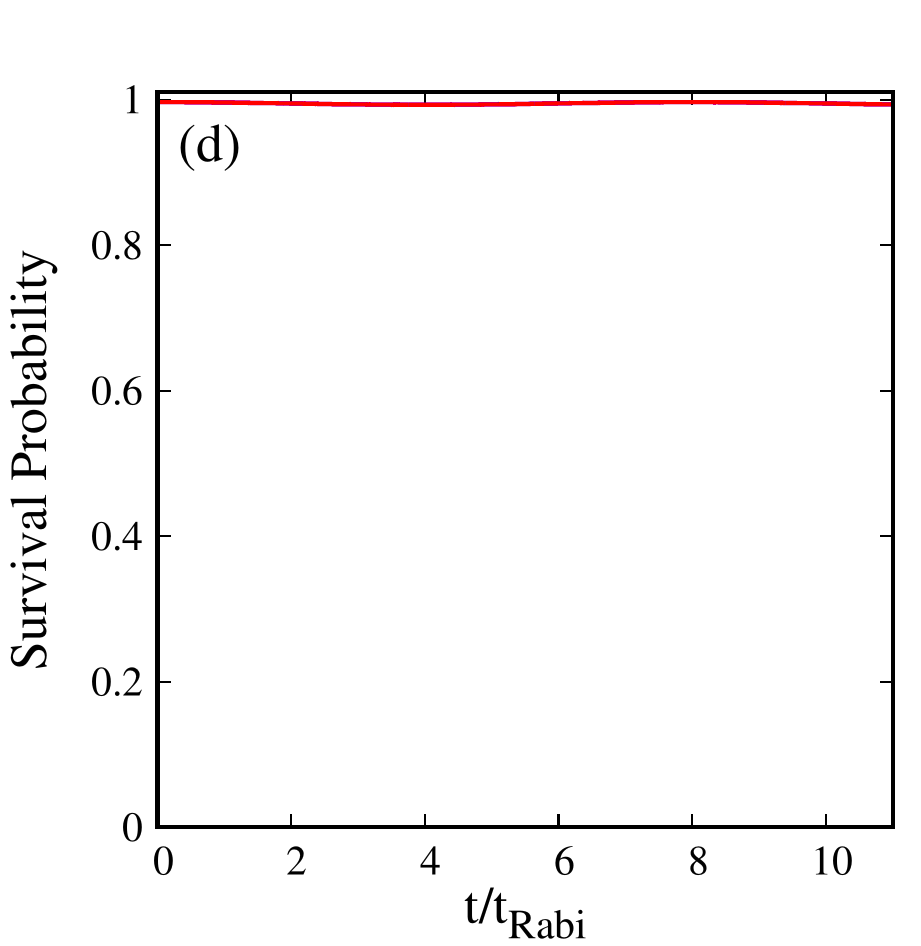}
\caption{Time-dependent survival probability characterizing the Josephson dynamics in position space in a rotating double well potential with interaction parameter $\Lambda=0.04$. Panel ($a$) corresponds to the survival probability for $\Omega=0$, panel ($b$) corresponds to survival probability for $\Omega=0.2$, panel ($c$) for $\Omega=0.3$, and panel ($d$) for $\Omega=0.5$. The blue colored line depicts the mean-field dynamics, and the red colored one represents the many-body dynamics. All the quantities are dimensionless.} 
\label{double_well}
 \end{figure*} 

\section{ Rotating a double-well trap and self-trapping}
 One area that remains largely unexplored is the Josephson dynamics of rotating bosons in a generic double-well potential. Here, the double-well dynamics is approached as a preparatory problem. We proceed further with our investigation on the basis of the findings of the rotating double-well dynamics. We choose $\alpha=1$ in the double-well potential in Eq. \eqref{eq:one}, which corresponds to a double well with barrier height $V_0=1.5$.
We examine the dynamics of rotating bosons in this double-well potential at both the mean-field and many-body levels of theory.
We compare the many-body results computed using the MCTDHB approach with the corresponding mean-field counterparts in order to emphasise the many-body facets of the dynamics.

Let us discuss the behavior of the survival probability that quantifies the dynamics of bosons in position space at different rotation frequencies $\Omega$. At $\Omega=0$, the usual tunneling dynamics of interacting bosons is observed at both the mean-field and many-body levels, see Fig.~\ref{double_well}(a). In the short-time dynamics, we find that the mean-field and many-body survival probabilities completely overlap with the presence of Rabi-like oscillations. This demonstrates that mean-field theory can adequately describe the dynamics at short time scales for finite interaction strength. The discrepancy between the mean-field and many-body dynamics becomes prominent in the long-time dynamics. As time progresses, the many-body dynamics exhibits incomplete tunneling of the bosons which is reflected in the gradual decay in the oscillation amplitude. These features indicate the development of many-body correlations. In contrast, such amplitude decay is absent at the mean-field level. 

It is apparent from Fig.~\ref{double_well}(b)-(c) that the tunneling period increases with the rotation frequency $\Omega$. In this context, rotation introduces an additional barrier on top of the existing barrier of the double well. Consequently, as $\Omega$ increases, the tunneling period becomes longer. In the supplemental material [see Sec. S3.B], we have provided a detailed discussion on how the effective barrier height increases with $\Omega$.

 Finally, for the critical rotation frequency $\Omega_c=0.5$, the tunneling collapses completely; Fig.~\ref{double_well}(d). Hence, the values of the mean-field and many-body survival probability stay at $100\%$ at all times, thus signaling the onset of complete self-trapping. In this context, all the bosons become localized in the left well as the value of $\Omega$ increases. The large centrifugal barrier associated with strong rotation prevents the bosons from tunneling between the wells. Moreover, we also note that the mean-field and many-body dynamics have a similar period of oscillation irrespective of the increment in $\Omega$ until $\Omega_c$ due to the treatment of a weakly interacting condensate. 
 
 In summary, we can conclude that the real-space dynamics of bosons in a rotating double-well reveals that rotation gradually slows down the dynamics with the increase of rotation frequency. At the critical rotation frequency, $\Omega_c$, tunneling is suppressed due to the increased effective centrifugal barrier. The corresponding momentum density, discussed in the supplemental material (in Sec.~S2), shows no signature of the emergence of a momentum-space Josephson junction.

\section{Protocol to generate a momentum-space Josephson junction}
Thus far, we have observed that the rapid rotation eliminates Josephson tunneling in the double-well potential, which results in self-trapping. To overcome this problem, we provide a protocol to host the Josephson dynamics in the rotating frame for sufficiently large rotation. We also show how to generate a momentum-space Josephson junction with rotation by altering the trap geometry to reduce the additional barrier. The effect of the centrifugal barrier caused by rotation in the double-well potential is compensated in this double-well-based protocol as \cite{gamma}
\begin{eqnarray}
\hat{V}_{C}(\mathbf{r}) &=& \hat{V}(\mathbf{r})+\frac{1}{2}\Omega^2 \mathbf{r}^2.
\label{eq:two}
\end{eqnarray}
Here, $\hat{V}(\mathbf{r})$ is the double-well potential used in this work, which is defined by Equation~\eqref{eq:one}. 

\begin{figure*}[htbp]
 \centering
\includegraphics[width=0.3\textwidth]{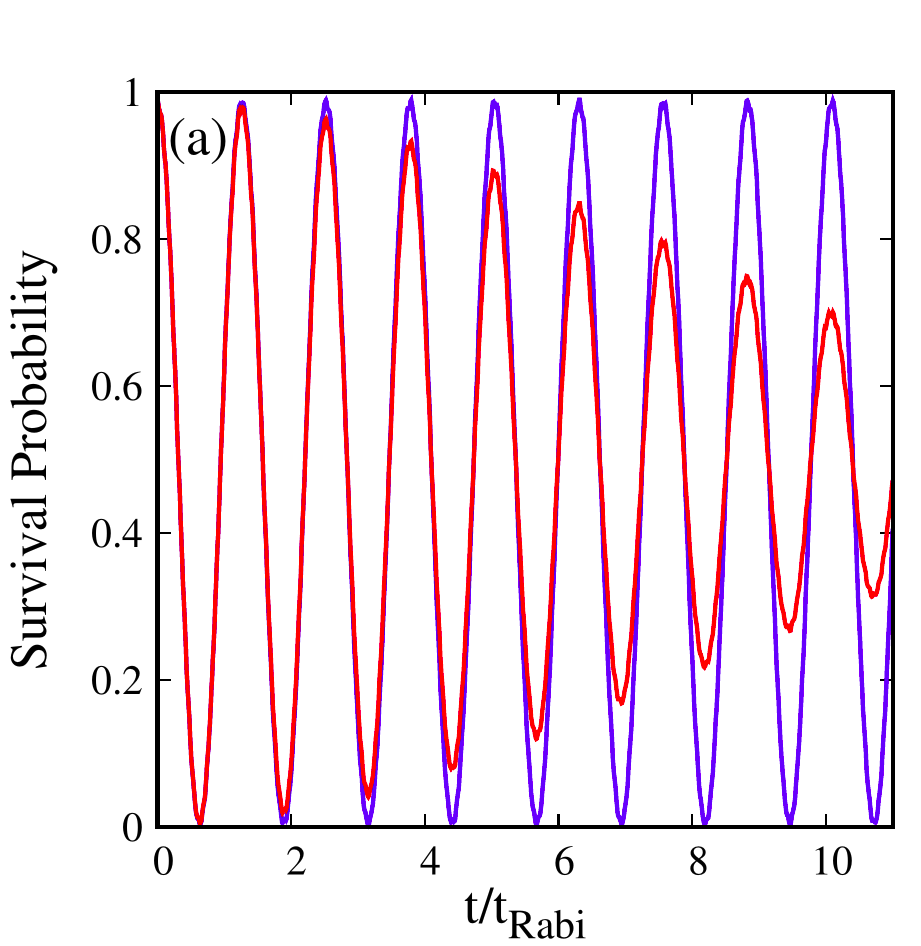}
\includegraphics[width=0.3\textwidth]{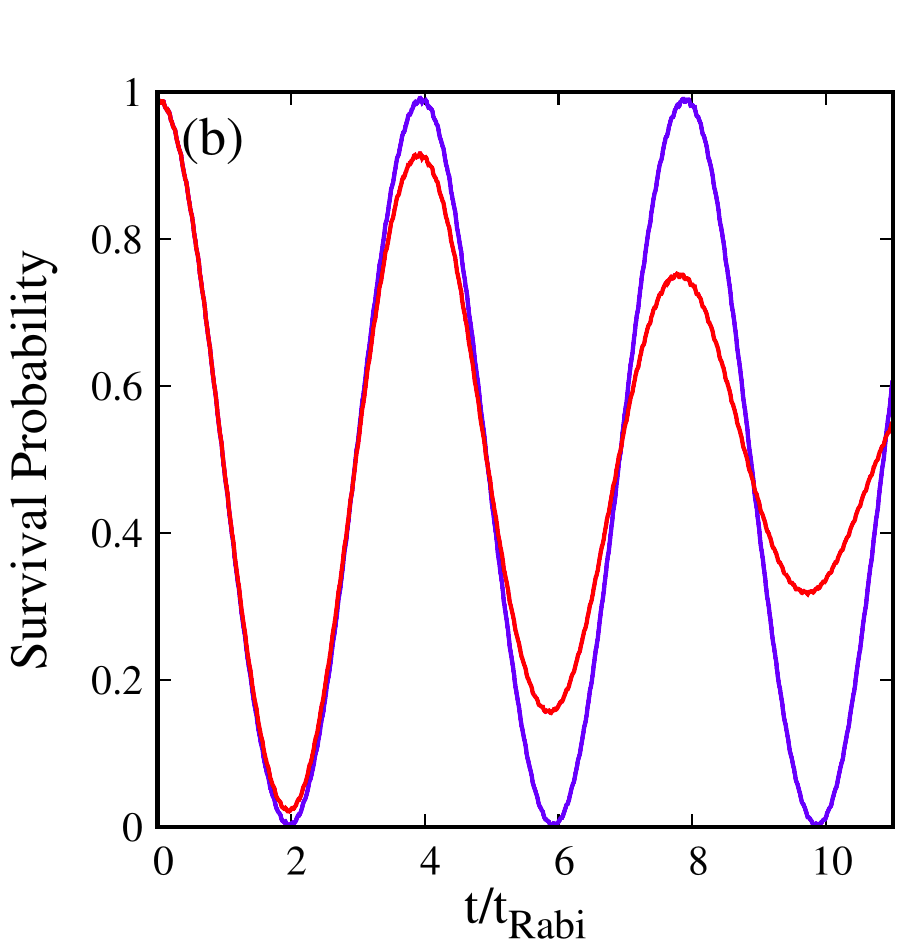}
\includegraphics[width=0.3\textwidth]{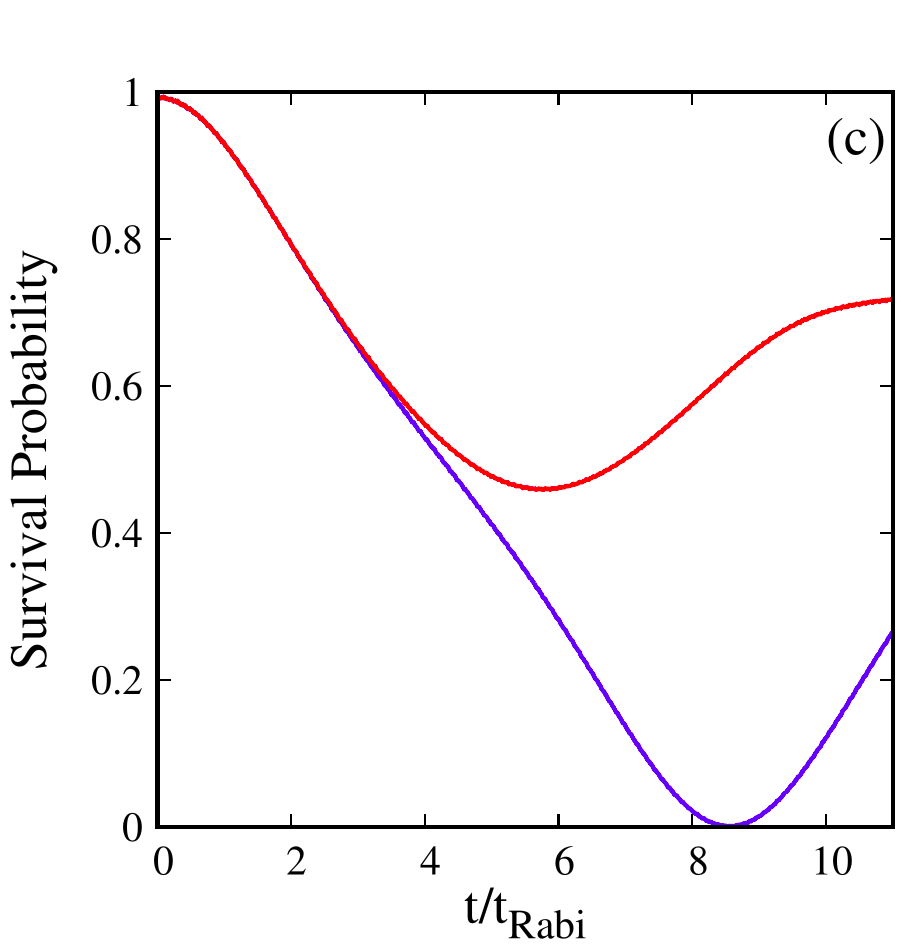}
 \caption{Dynamics of the survival probability characterizing the position-space Josephson dynamics are shown for the protocol to generate a momentum-space Josephson junction at three different rotation frequencies $\Omega$ for the interaction parameter $\Lambda=0.04$. Panel (a) corresponds to $\Omega=0.2$, panel (b) exhibits the results for $\Omega=0.5$, and panel (c) showcases the dynamics for $\Omega=0.7$. The blue-colored line depicts the mean-field dynamics, and the red-colored one represents the many-body dynamics. Beyond the critical rotation frequency of the double-well potential without the centrifugal barrier ($\Omega_c=0.5$), the Josephson dynamics is prominent for this protocol. All the quantities are dimensionless.}\label{surv_prob_pr_1}
 \end{figure*}  

At slow rotations ($\Omega=0.2$), the Josephson dynamics is comparable to that of a non-rotating double well. The mean-field and many-body survival probabilities coincide in the short-time dynamics. In long-time dynamics, the mean-field survival probability retains oscillations with constant amplitude, thereby referring to the complete tunneling of the densities; Fig.~\ref{surv_prob_pr_1}(a). However, the many-body dynamics shows damped oscillations, which is a signature of the development of quantum correlations in the condensate. 
At $\Omega=0.5$, which marks the rotation frequency that completely suppresses dynamics in a rotating double-well potential ($\Omega_c$), Josephson dynamics is still prominent in this protocol. The dynamics exhibit complete oscillations and a longer tunneling period, as shown in Fig. ~\ref{surv_prob_pr_1}(b). 
For a notably faster rotation ($\Omega=0.7$), the Josephson tunneling is still prominent with an enlarged tunneling period; see Fig.~\ref{surv_prob_pr_1}(c). Here, basically, the superposition of the extra-harmonic feed in the double-well potential in Equation \eqref{eq:two} squeezes the two minima of the combined potential, thereby aiding the Josephson tunneling in this protocol. 
\begin{figure}[]
 \centering
\includegraphics[width=0.7\textwidth]{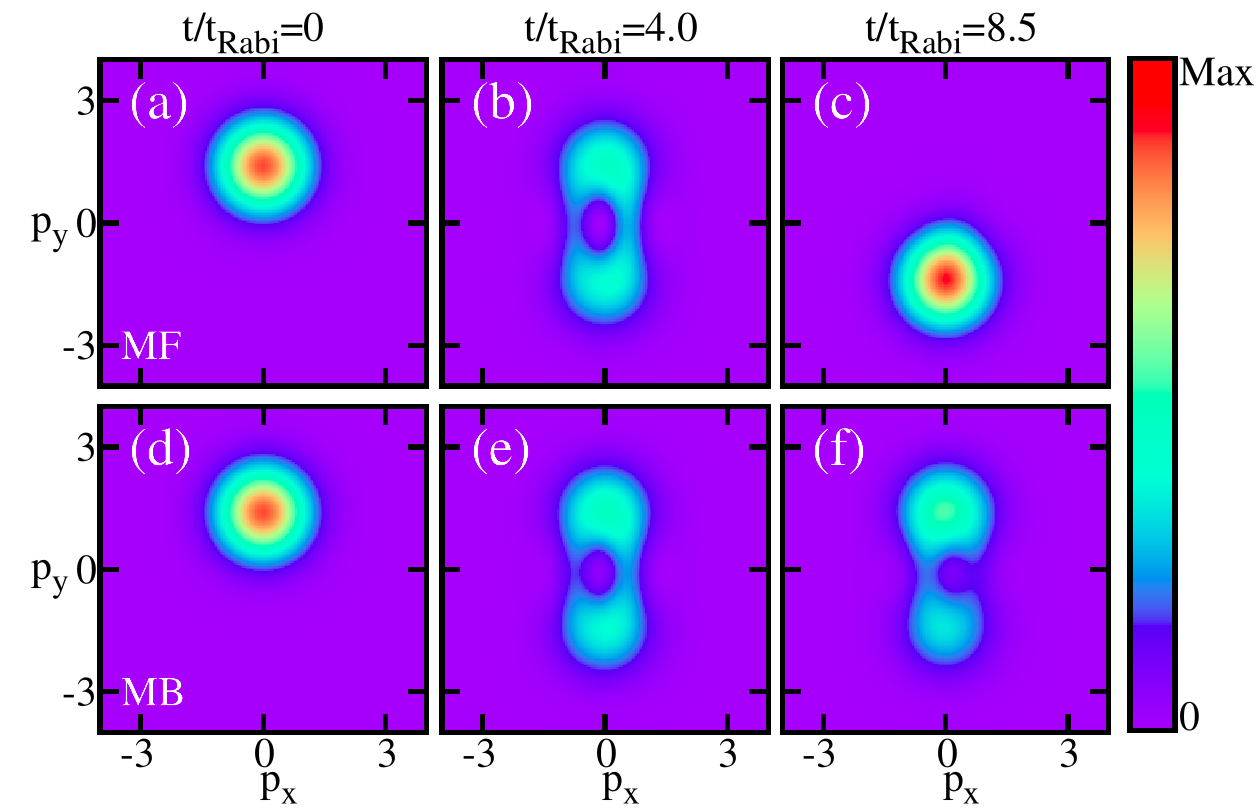}
\caption{Time snapshots of the momentum density per particle, which hallmark the emergence of the momentum-space Josephson dynamics for the double-well-based protocol for a specific rotation frequency $\Omega=0.7$ and interaction parameter $\Lambda=0.04$. Panels (a)-(c) represent the mean-field (MF) momentum densities per particle, and panels (d)-(f) correspond to the many-body (MB) momentum densities per particle. To capture the momentum-space Josephson tunneling, we have opted for three time frames, namely $t_1=0$, $t_2=4.0$, and $t_3=8.5$. All the quantities are dimensionless.}\label{mom_prop_1}
\end{figure}


The time snapshots of the momentum density profile for fast rotation ($\Omega=0.7$) are captured in Fig.~\ref{mom_prop_1}. In order to investigate the dynamics in momentum space, three time frames, $t_1=0$, $t_2=4t_{\text{Rabi}}$, and $t_3=8.5t_{\text{Rabi}}$, which represent the three stages of the first half of the tunneling period of the mean-field dynamics in real space, are used. $t_1$ refers to the initial time when all bosons are in the left well, and $t_2$ corresponds to the instant of time during tunneling when the two wells have the same occupancy. Similarly, $t_3$ corresponds to the instant of time when all the bosons have tunneled towards the right well. The full course of tunneling of the mean-field densities along the transverse direction of the momentum space is evident in Fig. \ref{mom_prop_1}(a)-(c). 
The many-body position-space dynamics exhibits incomplete tunneling, as reflected in the survival probability at $\Omega=0.7$ [Fig.~\ref{surv_prob_pr_1}(c)]. Correspondingly, the momentum-density dynamics also shows incomplete oscillations [Fig.~\ref{mom_prop_1}(d)-(f)], in contrast to the complete tunneling observed at the mean-field level. 
Hence, a sufficiently high rotation produces well-localized wavepackets in the momentum space, which renders the
onset of Josephson dynamics along the transverse direction $p_y$ in the momentum space. A slight revival of tunneling is also marked in the many-body dynamics [compare Figs. \ref{mom_prop_1} (e) $\&$ (f)]. 
The complete simulations of the Josephson dynamics in both position and momentum spaces are available in Sec.~S7 of the supplemental material. For the momentum-space analysis, the computational box is enlarged to $[-20,20)\times[-20,20)$ and discretized using a $128\times128$ grid points. This configuration maintains the same spatial grid resolution as in the position-space calculations while providing a finer momentum-space discretization, as the momentum grid spacing is given by $\Delta p = 2\pi/L$, where $L$ denotes the box length.

As the double well rotates in the clockwise direction, the system experiences a kick in the $p_y$ direction. The left well is pushed up, getting a positive kick in $p_y$ ($+p_y$ direction), and the right well is pushed down by experiencing a negative kick in $p_y$ ($-p_y$ direction). This affects the expectation value of the momentum, and $\langle p_y\rangle$ has two values now; one is around a positive kick and the other is around a negative kick. This signifies the emergence of the effective double well in the momentum space, which indicates the emergence of a momentum-space junction. Thus, the initial expectation value of $p_y$ picks up from zero to a finite positive value with an increase in rotation frequency $\Omega$.

\begin{figure*}[hptb]
\centering
 \includegraphics[width=0.35\textwidth]{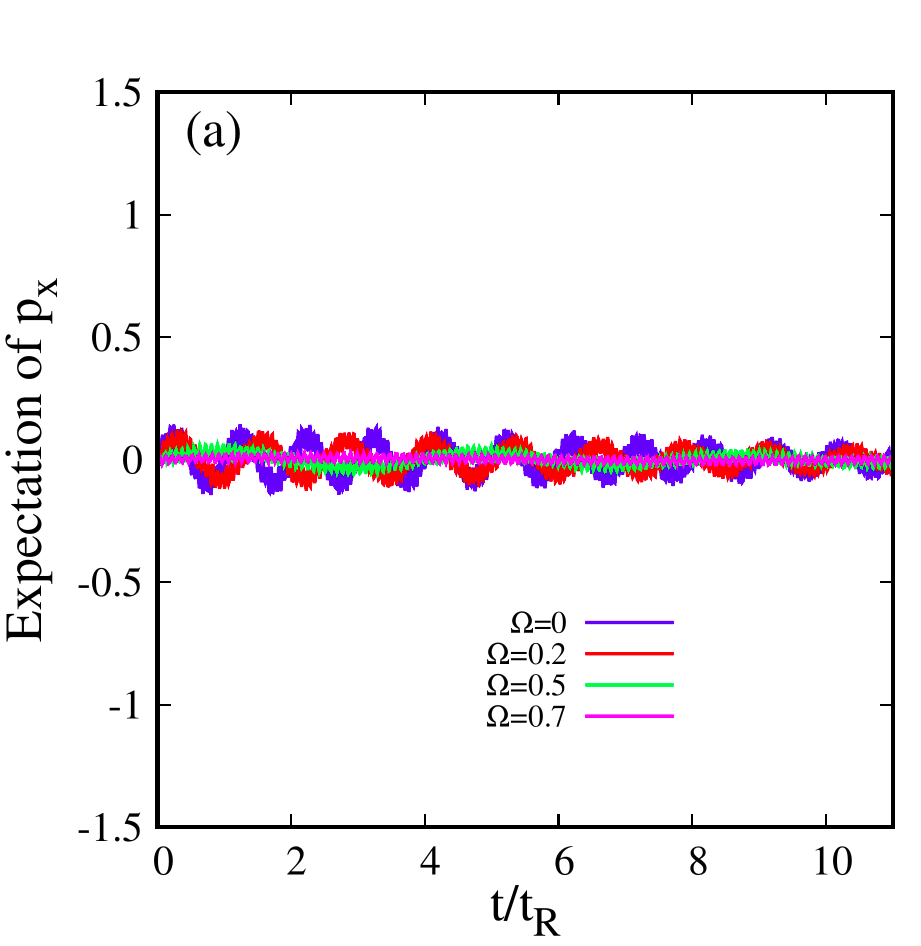}
 \includegraphics[width=0.35\textwidth]{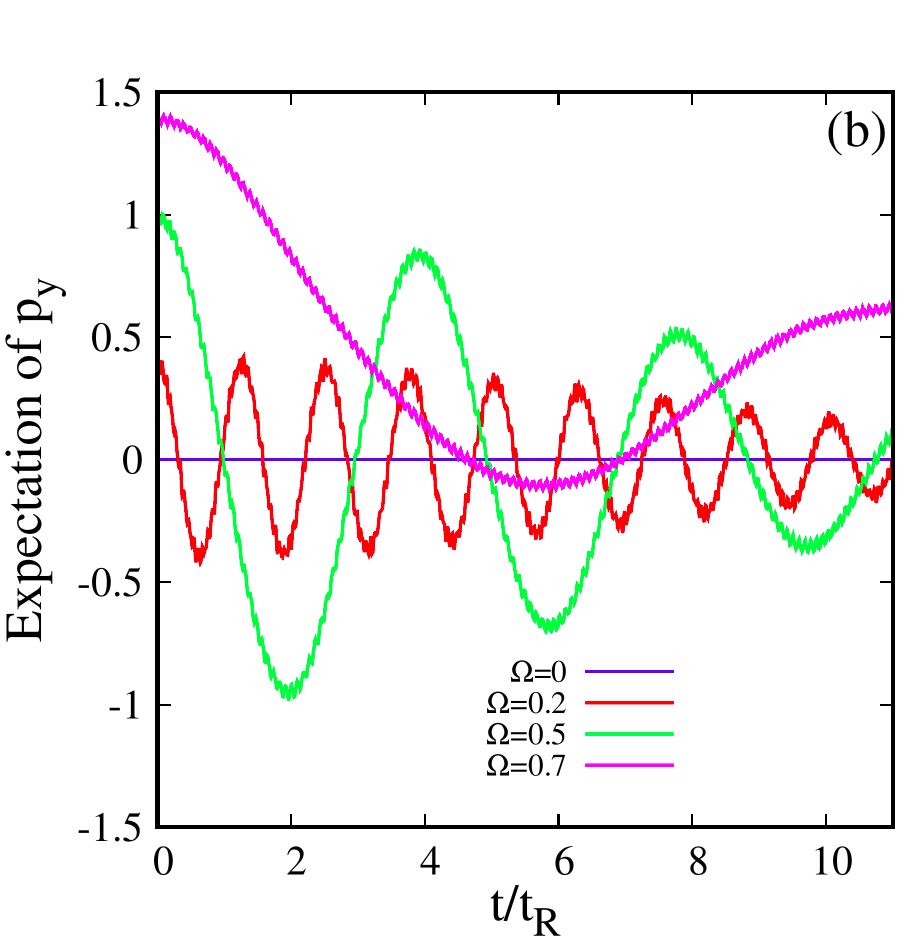}
\caption{The time evolution of the expectation values of linear momentum, $\langle p_x \rangle$ and $\langle p_y \rangle$, for several rotation frequencies $\Omega$ for a double-well-based protocol involving $N=10$ bosons and interaction parameter $\Lambda=0.04$. It is evident from (a) that as $\Omega$ increases, $\langle p_x \rangle$ approaches $0$. In contrast, $\langle p_y \rangle$ rises from zero to non-zero values as $\Omega$ increases in (b). All quantities mentioned are dimensionless.} \label{momentum_exp}
\end{figure*}

Fig. \ref{momentum_exp} shows the time evolution of expectation values of the linear momentum $\langle p_x \rangle$ and $\langle p_y \rangle$ both in the x and y directions. In the case of $\langle p_x \rangle$, it is evident that without rotation ($\Omega=0$), the expectation value of $\langle p_x \rangle$ oscillates between $\langle p_x \rangle \approx \pm 0.14$. As we feed rotation into the condensate, the value of $\langle p_x \rangle$ reduces significantly, and it slightly fluctuates around $\langle p_x \rangle \approx 0$. Basically, strong rotation leads to $ \langle p_x \rangle \rightarrow0$, although there is significant dynamics in position space. This is consistent with the absence of a velocity kick along the x-direction, with the orbitals remaining essentially Gaussian-shaped and localized around $p_x=0$.
The behavior of $\langle p_y \rangle$ is markedly different. As can be seen from Fig. \ref{momentum_exp}(b), in the case of no rotation $(\Omega=0)$, $\langle p_y \rangle=0$ irrespective of time. However, in the rotating frame ($\Omega=0.5$), a significant amount of momentum is pumped into the system, and the expectation value of $p_y$ oscillates about $\pm 1$. The value of $\langle p_y \rangle$ increases with a further increase in $\Omega$ [see for $\Omega=0.7$].
Thus, rotation predominantly generates momentum dynamics along the transverse $y$ direction, while suppressing the expectation value of the momentum along $x$. This behavior provides further insight into the momentum-space dynamics of the rotating Josephson junction.

In summary, we need the rotating frame to induce a momentum kick in the $+p_y$ and $-p_y$ directions. This mechanism is responsible for imprinting an effective double well in momentum space. 
For scalar bosons or a single-component Bose-Einstein condensate, the emergent momentum-space Josephson junction has not been previously documented, as far as we know, in the literature. This protocol not only facilitates Josephson dynamics along the longitudinal spatial direction for strong rotations but also establishes a Josephson junction in momentum space along the transverse direction.

\section{Rotation-generated Josephson junctions without a barrier}
 As an extension, it would be interesting to examine the effects of rotation on the condensate in the case where the double-well barrier is entirely removed [when $\alpha=0$ in Equation \eqref{eq:one}]. 
\begin{figure*}[hptb]
 \centering
\includegraphics[width=0.3\textwidth]{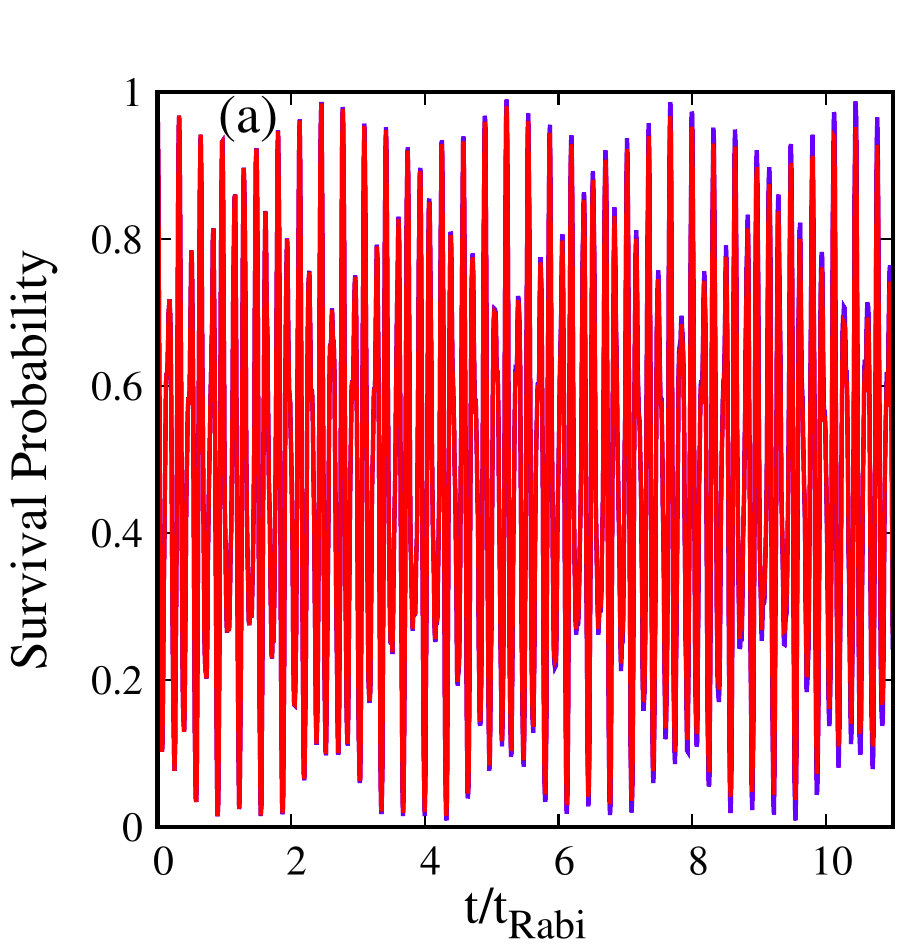}
\includegraphics[width=0.3\textwidth]{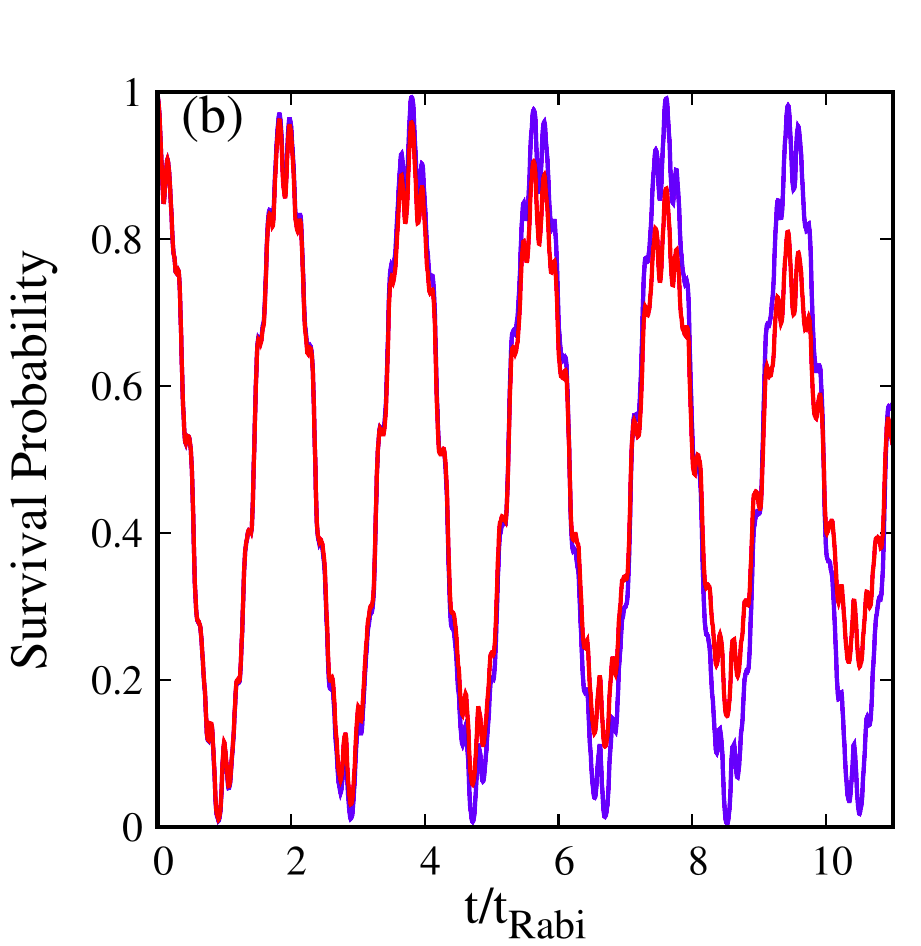}
\includegraphics[width=0.3\textwidth]{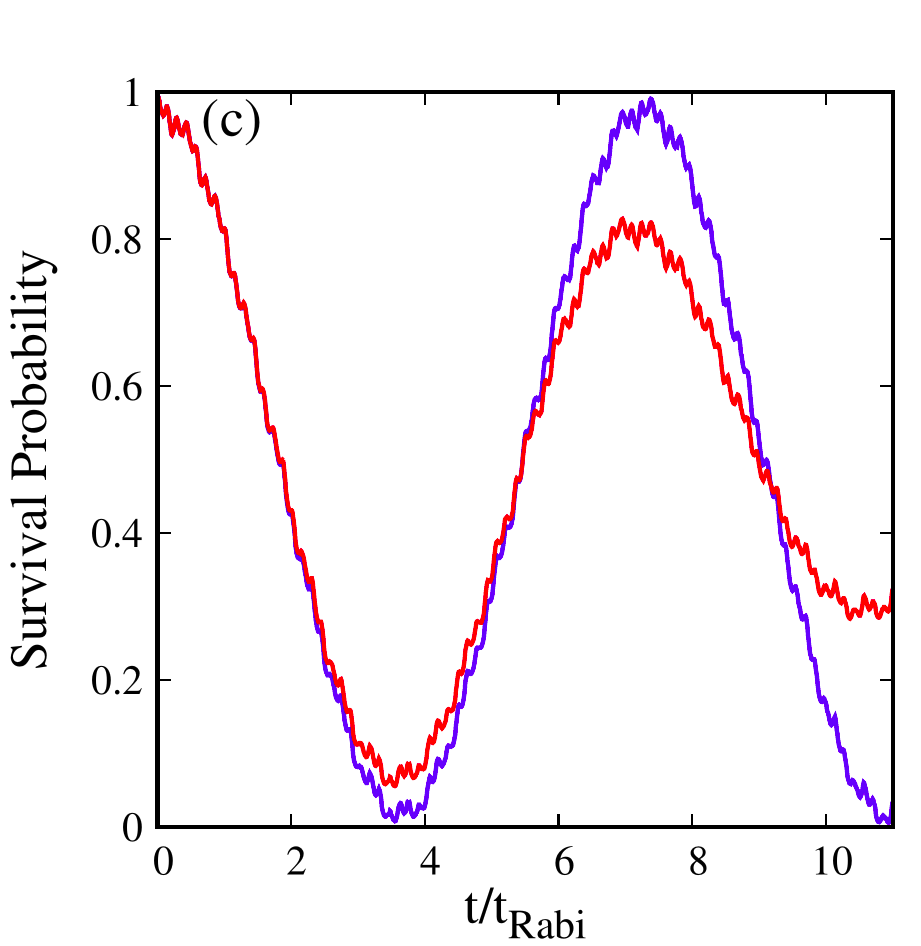}
 \caption{Dynamics of the survival probability characterizing the Josephson dynamics in position space for various rotation frequencies $\Omega$ in a barrierless trap with interaction parameter $\Lambda=0.04$. Panel (a) corresponds to $\Omega=0$, panel (b) represents $\Omega=0.45$, and panel (c) gives the results for $\Omega=0.5$. The blue-colored line depicts the mean-field dynamics and the red-colored one represents the many-body dynamics. All the quantities are dimensionless.}\label{surv_prob_pr_2}
 \end{figure*}

Fig.~\ref{surv_prob_pr_2} depicts the mean-field and many-body dynamics of the survival probability in the barrierless trap. In the absence of rotation, the finite interaction and dispersion of the bosons cause the density to wobble in the single-well potential. As rotation is introduced in the condensate, the tunneling of bosons becomes slightly visible. Now, the Josephson dynamics is prominent at the critical rotation frequency $\Omega_c=0.5$ of the double-well potential. However, for fast rotations, $\Omega>\Omega_c$ results in the emergence of self-trapping. In Sec.~S6.A of the supplemental material, we discuss a comparative analysis of survival probabilities for strong rotation frequencies to detect the onset of self-trapping.

\begin{figure}[htbp]
 \centering
\includegraphics[width=0.7\textwidth]{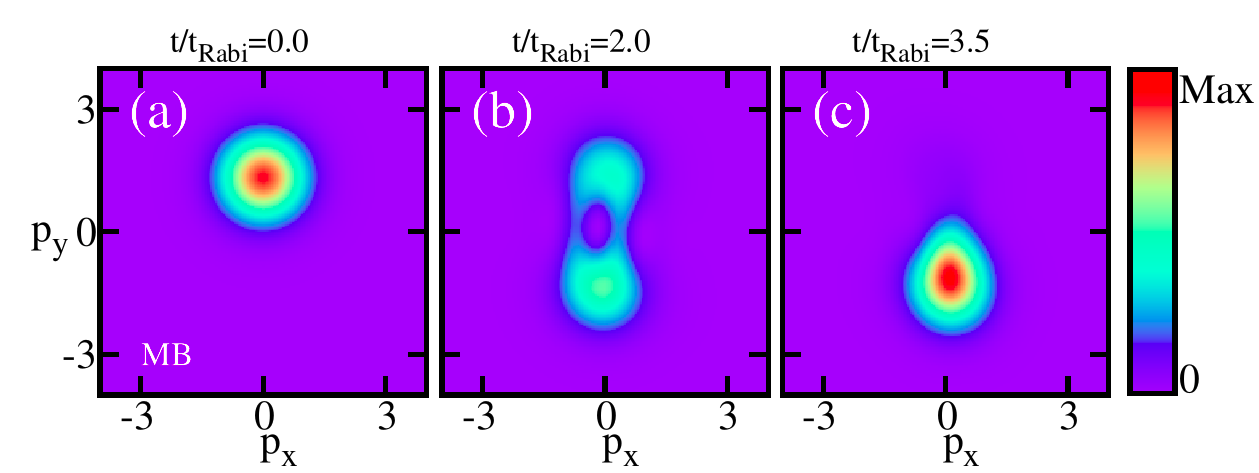}
\caption{Time snapshots of the momentum density per particle that mark the emergence of the momentum-space Josephson dynamics in a single-well trap for rotation frequency $\Omega=0.5$ with interaction parameter $\Lambda=0.04$. Panels (a)-(c) represent the many-body (MB) momentum densities per particle. The mean-field densities exhibit similar dynamics and hence are not displayed. To capture the momentum-space Josephson dynamics, three time frames are chosen, namely $t_1=0$, $t_2=2.0$, and $t_3=3.5$. The significance of these three time frames is elaborated in the main text. All the quantities are dimensionless.}\label{mom_prop_2}
\end{figure}

Several time snapshots of the momentum density profile in the rotating single-well potential are shown in Fig. (\ref{mom_prop_2}). Interestingly, this study also demonstrates the emergence of a momentum-space Josephson junction in the transverse direction at sufficiently large rotation, $\Omega=0.5$. Here, the time frame $t_1=0$ is the initial time when all bosons occupy the left well, $t_2=2.0 t_{\text{Rabi}}$ corresponds to the time when each well is essentially equally populated, and $t_3=3.5 t_{\text{Rabi}}$ presents the instant of time when all bosons tunnel to the right well. These three time frames correspond to the three stages of the first half of the tunneling period of the survival probability. 
It is also confirmed that the mean-field and many-body dynamics of the momentum densities are similar here. The complete density evolution is provided in Sec. S7 of the supplemental material.

In summary, our work has paved the way for engineering Josephson junctions in position and momentum spaces in the rotating frame by modifying the geometry of a double-well potential. For the first time, rotation is observed to create momentum-space Josephson junctions with scalar bosons, which has so far been achieved only for spinful bosons with spin-orbit coupling.

\section{Concluding remarks}
In a rotating frame, this work offers a method to create a position-space Josephson junction in the longitudinal direction as well as a momentum-space Josephson junction along the transverse direction for scalar or single-component bosons. Here, we report protocols to generate rotation-mediated bosonic Josephson dynamics in both position and momentum spaces by manipulating the geometry of a double-well trap and the rotation frequencies. These momentum-space Josephson junctions are expected to have important applications in fabricating novel quantum mechanical devices and in quantum sensors. Due to the higher degree of tunability of the tunneling of the condensate and their stability in a momentum-space Josephson junction, they are significant for prospective applications in fabricating novel quantum mechanical devices. Our finding may inspire further experimental and theoretical investigation of momentum-space Josephson junctions for single-component bosons. Another promising direction is to explore distinct features of Josephson dynamics by exploiting momentum-space tunneling as an additional degree of freedom.

In the present work, a synthetic gauge field is considered to introduce rotation into the condensate (see the supplemental material). In this context, the rotation can be considered as a particular case of the synthetic gauge field. Thus, the possibility of creating a momentum-space Josephson junction using a synthetic gauge field for scalar bosons would be an intriguing avenue for future research. Finally, vortex nucleation is an important feature of a rotating condensate, and the nucleation of quantized vortices is a phenomenon characterised by intense rotations and significantly highly interacting systems. In this context, it would be fascinating to extend our system to the regime corresponding to vortex nucleation.

\section{Acknowledgments}
This work was supported by the Israel Science Foundation (Grant no. 1516/19). Computation time on the High Performance Computing Center Stuttgart (HLRS) is gratefully acknowledged. The authors acknowledge Dr. Axel U. J. Lode for insightful feedback.

\bibliographystyle{apsrev4-2-title} 
\bibliography{manuscript}

 \end{document}